\documentclass[%
 reprint,
superscriptaddress,
 amsmath, amssymb,aps,
floatfix,
]{revtex4-1}

\usepackage{graphicx}
\usepackage{dcolumn}
\usepackage{bm}
\usepackage[colorlinks=true,citecolor=blue,linkcolor=magenta]{hyperref}
\usepackage[utf8]{inputenc}
\usepackage{url}
\usepackage[normalem]{ulem}
\usepackage{upgreek}
\usepackage[mathlines]{lineno}
\usepackage{amsmath}

\makeatletter
\newcommand*{\balancecolsandclearpage}{%
  \close@column@grid
  \twocolumngrid
}

\begin{document}


\title{Monolithic Kerr and electro-optic hybrid microcombs}
\author{Zheng Gong}
\affiliation{Department of Electrical Engineering, Yale University, New Haven, CT 06511, USA}

\author{Mohan Shen}
\affiliation{Department of Electrical Engineering, Yale University, New Haven, CT 06511, USA}

\author{Juanjuan Lu}
\affiliation{Department of Electrical Engineering, Yale University, New Haven, CT 06511, USA}

\author{Joshua B. Surya}
\affiliation{Department of Electrical Engineering, Yale University, New Haven, CT 06511, USA}

\author{Hong X. Tang}
\affiliation{Department of Electrical Engineering, Yale University, New Haven, CT 06511, USA}
\affiliation{Corresponding author: hong.tang@yale.edu}
\date{\today}
\renewcommand{\figurename}{Fig.}

\begin{abstract}
\vspace{12pt}
Advances in microresonator-based soliton generation promise chip-scale integration of optical frequency comb for applications spanning from time keeping to frequency synthesis. Miniaturized cavities harness Kerr nonlinearity and enable terahertz soliton repetition rates. However, such high repetition rates are not amenable to direct electronic detection. Here, we demonstrate hybrid Kerr and electro-optic microcombs using the lithium niobate thin film that exhibits both Kerr and Pockels nonlinearities. By interleaving the high-repetition-rate Kerr soliton comb with the low-repetition-rate electro-optic comb on the same waveguide, the wide Kerr soliton mode spacing is divided within a single chip, allowing for subsequent electronic detection and feedback control of the soliton repetition rate. Our work establishes an integrated electronic interface to Kerr solitons of terahertz repetition rates, paving the path towards chipscale optical-to-microwave frequency division and comb locking.
\end{abstract}

\maketitle

\setcounter{figure}{0} 
\renewcommand{\thefigure}{\textbf{\arabic{figure}}}
\renewcommand{\figurename}{\textbf{Fig}}

\section{Introduction}

Optical frequency combs are precision spectral rulers that have revolutionized frequency and time measurements \cite{Diddamunite}. Since early demonstrations of frequency combs with solid-state mode-locked lasers, persistent efforts have been made to advance the comb technology from laboratory concepts to field deployment \cite{Fortieryears}. An important progress is the soliton mode-locking in high-quality-factor Kerr microresonators \cite{Herr2013Temporal,Xue2015Mode-locked} which pushes comb systems towards integrated chips \cite{heterogenousCombdevice} for portable operation \cite{Stern2018Battery}. Due to the tight optical confinement in a microresonator, the Kerr ($\chi^{(3)}$) nonlinearity is enhanced, leading to sub-milliwatt comb initiation \cite{submilliwatts,submilliwatt2} as well as new avenues for studying soliton physics \cite{Yang2016Stokes,Cole2017Soliton,heteronuclear,photonicdimer}. The versatile waveguide dispersion control further permits broadband comb operation \cite{Li2017Stably,Pfeiffer2017Octave,LNnearoctave, midIR,AlNbeyondoctave,He_octavecomb}. To date, a multitude of nanophotonic material platforms \cite{Kippenberg2018Dissipative} have been developed to support soliton microcomb sources which have already enabled new capabilities in optical clock \cite{Newman2019Architecture,Drake2019Terahertz}, optical frequency synthesis \cite{synthesizer}, spectroscopy \cite{Dutt2018On-chip,spectroscopy2} and etc \cite{Kippenberg2018Dissipative}.

To further the integration of microcomb systems for scalable production and high volume applications, efforts have been made to integrate soliton initiation circuitry onto chip \cite{Raja2019Electrically,Hertzlasercomb,Stern2018Battery,heterogenousCombdevice}. Another key element is the detection and control of the soliton repetition rates \cite{offchip_eo,Yu2019Tuning,Drake2019Terahertz,VernierFrep}. Despite that Kerr solitons with microwave repetition rates are achievable \cite{RFRepcomb,RFRepnonchipcomb}, broadband self-referenceable solitons often exhibit much wider mode spacings in the terahertz range which are usually not amenable to direct photodetection \cite{Pfeiffer2017Octave,Li2017Stably,zhang2019terahertz}. Previous works divided such high soliton repetition rates by the use of a separate Kerr microcomb device \cite{synthesizer,VernierFrep} or off-chip electro-optic (EO) modulators \cite{offchip_eo,Yu2019Tuning,Drake2019Terahertz}. Under cascaded phase modulation, dense sidebands can be created between neighboring soliton comb lines to facilitate the electronic detection and stabilization of the soliton repetition rate \cite{offchip_eo}. 

Meanwhile, a flurry of studies have been devoted to exploiting the additional $\chi^{(2)}$ nonlinearity existing in non-centrosymmetric Kerr microcomb materials \cite{Jung:14,Wilson2019Integrated,He2019Self-starting}. Not only do they unveil novel nonlinear dynamics \cite{Bruch2020Pockels,vischerenkov}, they also bring about opportunities to extend Kerr microcomb system functionalities, such as phase-matched frequency doubling \cite{Lu2019Periodically,doulerLuo:18,Bruch201817000}, and EO modulation \cite{Wang2019Monolithic,Zhang2019Broadband,LNreview}.

Here, we demonstrate an integrated approach to detect and control the high repetition rate of Kerr solitons using a monolithic hybrid Kerr and EO microcombs circuit patterned on a single z-cut lithium niobate (LN) thin film. Owing to the coexistence of Kerr ($\chi^{(3)}$) and Pockels ($\chi^{(2)}$) nonlinearities on this emerging nanophotonic platform \cite{He2019Self-starting,Gong2019Soliton,okawachi2020chip,RamanYu,Wang2019Monolithic,Zhang2019Broadband,gong2020photonic}, Kerr soliton and EO combs can be produced simultaneously within a single photonic circuit. The fast solitons and slow EO pulses converge onto a common output waveguide and lead to optical-heterodyning between the two pulse trains. In this way, a soliton repetition rate of 334\,GHz is divided down to a low frequency RF beatnote which can be processed electronically and monitored in real time. We then show the dependence of soliton repetition rate on pump frequency and pump power, and further stabilize it to an external microwave reference. Our work establishes an integrated interface connecting terahertz-rate Kerr solitons and readily detectable microwaves, providing a path for on-chip photonic frequency division as well as comb locking. 

\begin{figure*}[!t]
\centering
\includegraphics[scale=1]{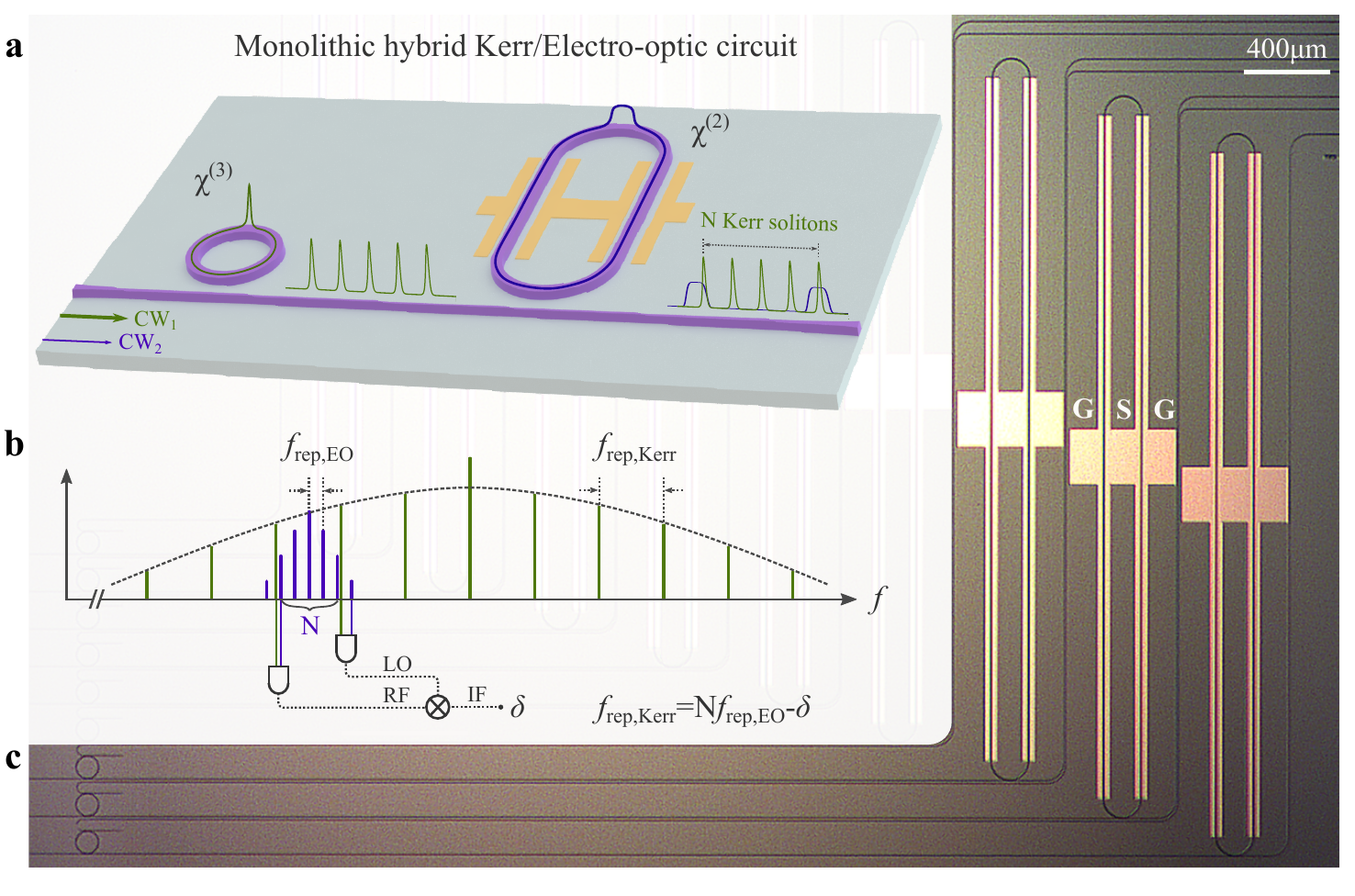}
\caption{\textbf{Experimental architecture for hybrid Kerr and EO microcombs integration}. \textbf{a} Schematic of a LN hybrid Kerr and EO circuit. Two c.w. pumps are used to separately generate high-repetition rate Kerr solitons (green) and low-repetition rate EO pulses (purple) from the two microrings. \textbf{b} The soliton comb (green) is spectrally interleaved with the dense EO comb (purple). Through successive detection and mixing of the hybrid combs heterodyne beat notes, $f_{\mathrm{rep,Kerr}}$ can be obtained. \textbf{c} Optical micrograph of an array of hybrid Kerr and EO circuits with varying bus-waveguide-microring coupling gaps on an area of 5-by-5\,$\mathrm{mm^{2}}$. The add-drop microring Kerr soliton generators are placed at the left side of the chip; the all-pass microring EO comb generators are set at the other side. S (G) denotes the signal (ground) electrode.}
\label{fig1}
\end{figure*}

\section{Results}
\noindent \textbf {Experimental scheme and on-chip hybrid Kerr and EO circuit.} Fig.\,\ref{fig1}a illustrates the schematic of a LN thin film hybrid Kerr and EO microcombs circuit. The Kerr soliton microcomb is generated by a c.w. telecom-band pump laser, while the EO comb is sourced by an auxiliary c.w. laser and driven by an RF reference synthesizer. When the RF frequency is tuned to match the EO microring FSR, an EO comb is formed and spectrally positioned in between adjacent soliton forming modes. The neighboring soliton and EO comb lines yield four low-frequency RF heterodyne beat notes. By subsequently mixing a selected pair of the beat notes, e.g., as shown in Fig.\,\ref{fig1}b, the Kerr soliton repetition rate ($f_{\mathrm{rep,Kerr}}$) is given by
\begin{equation}
f_\mathrm{rep,Kerr}=Nf_\mathrm{rep,EO}-\delta
\label{eq1}
\end{equation}
with $\mathrm{N}$ denoting the number of EO sidebands between the two soliton modes, $f_{\mathrm{rep,EO}}$ being the EO comb repetition rate and $\delta$ representing the frequency difference of the two selected beat notes (Fig.\,\ref{fig1}b). The involvement of the other RF heterodyne beat notes will be discussed in a later section. Note that the measurement of $f_{\mathrm{rep,Kerr}}$ is independent of the EO comb pump laser frequency and its frequency noise, which are cancelled in the coherent mixing process.

Our hybrid Kerr and EO circuits are monolithically fabricated on a millimeter-sized z-cut LN thin film (see Methods), as shown in Fig.\,\ref{fig1}c. The circuit consists of a microring Kerr soliton generator and a microring EO comb generator that are serially coupled to a common bus waveguide. The bus waveguide couples separate pump sources into the two microrings and delivers the resulting interleaved hybrid combs for $f_{\mathrm{rep,Kerr}}$ detection. A separate drop-port waveguide is added alongside the Kerr soliton microring for direct access of the soliton spectrum. This convenient circuit configuration requires only one pair of optical input and output ports to produce and access the hybrid Kerr soliton and EO combs for $f_{\mathrm{rep,Kerr}}$ measurement. The use of the auxiliary laser to generate the EO comb can be eliminated, for example, by electro-optically aligning an EO microring resonance to a soliton comb line.

\begin{figure}[t!]
\centering
\includegraphics[width=\linewidth]{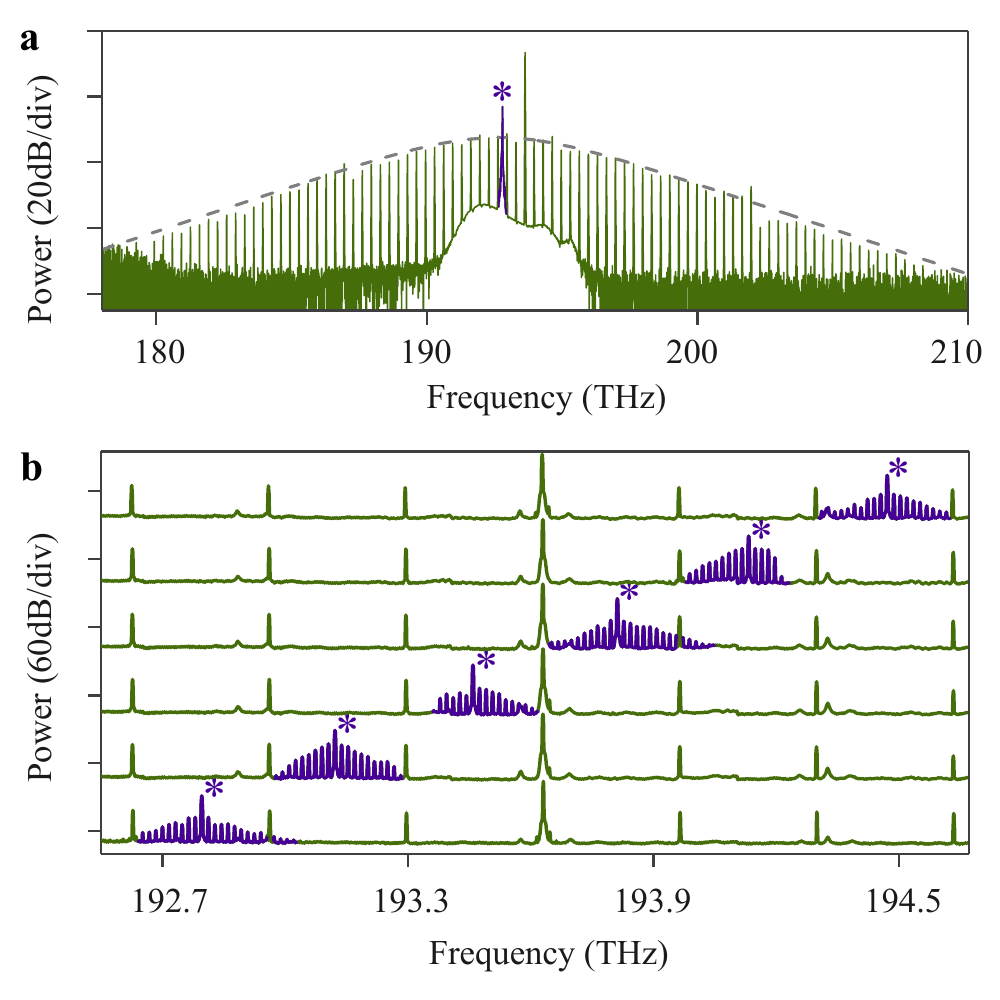}
\caption{\textbf{Hybrid Kerr and EO microcombs}. \textbf{a} The hybrid Kerr and EO combs spectrum output from the bus waveguide, measured with a resolution bandwidth (RBW) of 6.25\,GHz. The single-soliton spectrum (green) follows a $sech^{2}$-profile, exhibiting a full width at half-maximum (FWHM) of 5.6\,THz. The EO comb is colored purple with its pump marked by the asterisk. \textbf{b} Zoom-in view of hybrid microcombs spectra with the EO combs pumped at different wavelengths, measured at a 330-MHz RBW.}
\label{fig2}
\end{figure}

\noindent \textbf{Hybrid Kerr soliton and EO microcombs generation.} The Kerr soliton microring is geometrically engineered to exhibit anomalous dispersion for the $\mathrm{TE}_{00}$ modes at telecom wavelengths (see Methods). The microring coupling rates to external waveguides are adjusted to favor soliton formation over the stimulated Raman scattering (SRS) \cite{gong2020photonic,LNnearoctave} (see Methods). A soliton comb can be generated under an on-chip pump power of 50\,mW at 1549.5\,nm, featuring a 334-GHz mode spacing. By leveraging the photorefractive effect in z-cut LN microrings, soliton mode-locking is triggered by slowly scanning the red-detuned c.w. pump wavelength \cite{He2019Self-starting, Gong2018High-fidelity}.

The EO comb, generated in a racetrack microring-based EO modulator, harnesses resonantly enhanced EO modulation strength \cite{Zhang2019Broadband}. Favorable phase-matching condition for the $\mathrm{TE}_{00}$ EO comb is achieved by tailoring the microring geometries (see Methods). Here, in-plane electric fields are applied across the microring to modulate the $\mathrm{TE}_{00}$ modes through the LN EO tensor component $r_{22}$ (5.4\,pm/V \cite{yariv2007photonics}). The gap between the microring and the in-plane GSG gold electrodes is set to be 2\,${\mu}$m to provide strong RF drive fields while not attenuating the intracavity optical fields. 

Similar to Kerr soliton generation, the auxiliary c.w. pump can be steadily launched into a $\mathrm{TE_{00}}$ resonance of the EO microring from the red-detuned region. In experiments, a 1.5-mW on-chip optical pump and an RF drive with an approximately 14-V peak amplitude were used to generate the EO combs (Fig.\,\ref{fig2}). The pump power was set below the thresholds of Kerr four-wave mixing and SRS which otherwise could disturb the EO comb formation. The RF modulation field was provided by a microwave synthesizer running at a frequency close to the microring FSR ($\sim$\,16\,GHz). The measured microcomb spectra under different EO pump wavelengths are shown in Fig.\,\ref{fig2}b. At certain wavelengths, mode-crossings between the EO comb $\mathrm{TE_{00}}$ modes and other cavity modes are present and causes the asymmetry of the comb profile (see Supplementary Note 1). By selecting a suitable pump wavelength, well-developed EO comb can be obtained to bridge the gap of adjacent soliton lines. In the following $f_{\mathrm{rep,Kerr}}$ measurement, we chose to pump the $\mathrm{TE}_{00}$ resonance at 1556\,nm. 

\begin{figure}[t!]
\centering
\includegraphics[width=\linewidth]{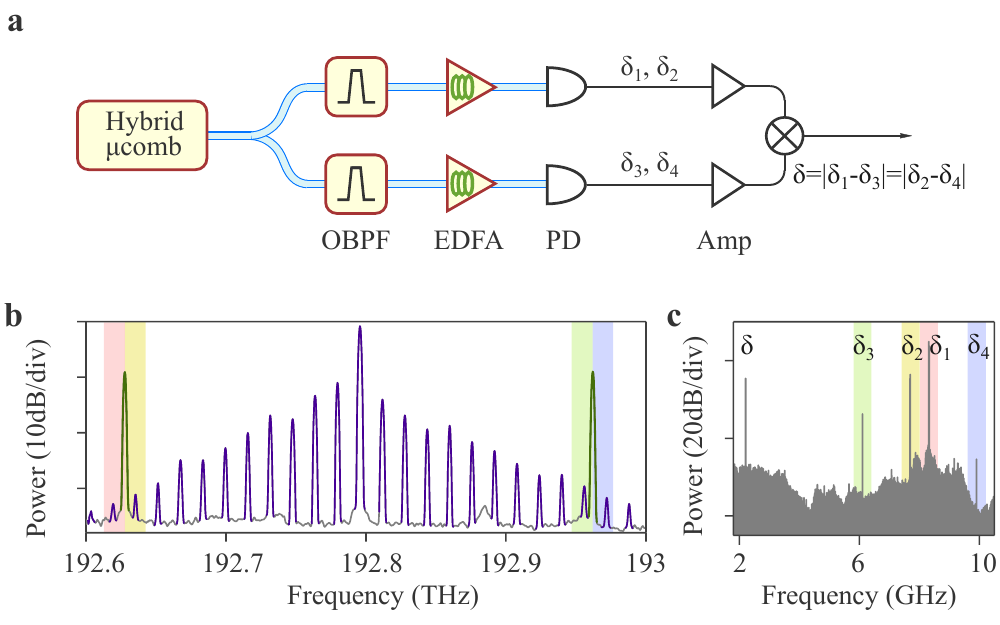}
\caption{\textbf{Electronic measurement of $f_{\mathrm{rep,Kerr}}$}. \textbf{a} Schematic for the measurement. EDFA, erbium-doped fiber amplifier; OBPF, optical band-pass filter; PD, photodetector; Amp, RF amplifiers. \textbf{b} Zoom-in view of the hybrid combs spectrum in Fig.\,\ref{fig2}(\textbf{a}), from which $f_{\mathrm{rep,Kerr}}$ is measured. The selected four pairs of comb lines for optical-heterodyne detection are highlighted. \textbf{c} The mixer output spectrum at a 50-kHz RBW. The four optical-heterodyne beat notes are highlighted in accordance with the heterodyning comb lines in (\textbf{b}). Their frequencies are labeled by $\delta_{\mathrm{i}}$ respectively.}
\label{fig3}
\end{figure}

The hybrid combs spectrum output from the bus waveguide (Fig.\,\ref{fig2}a) shows that the Kerr soliton comb spectral profile remains largely intact after passing through the EO microring. This is because the likelihood for soliton comb lines being aligned with EO microring resonances is low. Meanwhile, one can also obtain an unfiltered soliton spectrum from the drop-port of the soliton microring (see Supplementary Note 2). Here, the circuit allows both combs to exploit identical polarization state ($\mathrm{TE}_{00}$), facilitating the following heterodyne measurement.

\begin{figure*}[t]
\centering
\includegraphics[width=0.9\linewidth]{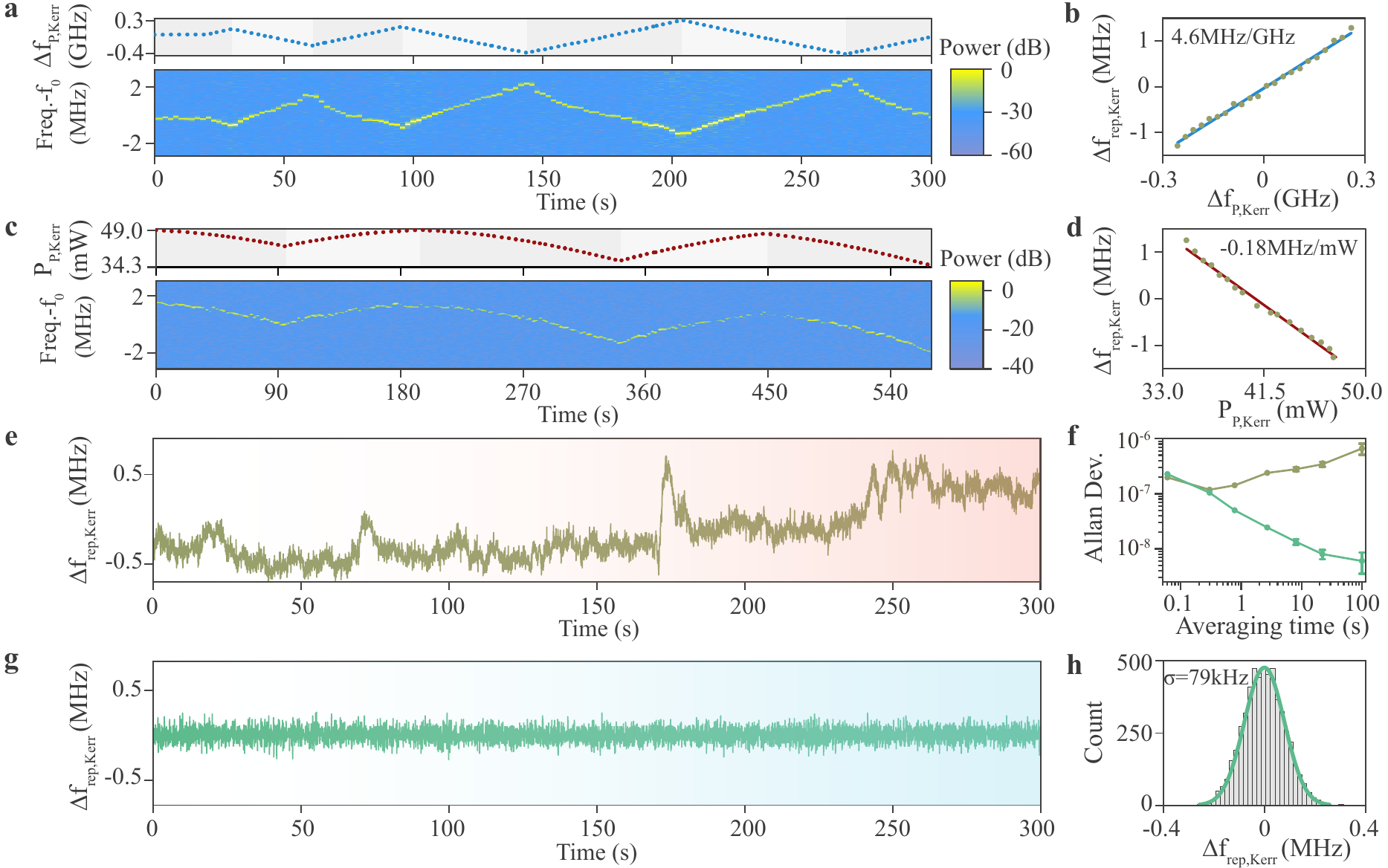}
\caption{\textbf{$f_{\mathrm{rep,Kerr}}$ control and stabilization}. \textbf{a} Kerr pump frequency tuning offset from 193.6\,THz (upper panel), and the recorded $\delta$ beat note spectra (lower panel) at a 40-KHz RBW. $\mathrm{f_{0}}$\,$=$\,2.2146\,GHz. \textbf{b} Dependence of $f_{\mathrm{rep,Kerr}}$ on the pump frequency extracted from (\textbf{a}). The measured values (dots) are linearly fitted (line). \textbf{c} On-chip Kerr pump power tuning (upper panel), and the recorded $\delta$ beat note spectra (lower panel) under the same RBW as in (\textbf{a}). $\mathrm{f_{0}}$\,$=$\,2.2126\,GHz. \textbf{d} Dependence of $f_{\mathrm{rep,Kerr}}$ on the on-chip pump power extracted from (\textbf{c}). \textbf{e} Frequency drift of the free-running $f_{\mathrm{rep,Kerr}}$. Each data point is measured with a 40-kHz RBW every 60\,ms. \textbf{f} Allan deviations of the free-running $f_{\mathrm{rep,Kerr}}$ (dark yellow) and the stabilized $f_{\mathrm{rep,Kerr}}$ (green). \textbf{g} Frequency drift of the stabilized $f_{\mathrm{rep,Kerr}}$ measured with the same settings as in (\textbf{e}). \textbf{h} Distribution of the frequency drift in (\textbf{g}), indicating a 79-kHz standard deviation and a 1-kHz standard error of the mean.}
\label{fig4}
\end{figure*}

\noindent \textbf{Detection and control of $f_{\mathrm{rep,Kerr}}$.} The experimental scheme for real-time electronically accessing $f_{\mathrm{rep,Kerr}}$ is illustrated in Fig.\,\ref{fig3}a. The outgoing hybrid microcombs are split into two paths to access different optical-heterodyne beat notes. Two tunable OBPFs are utilized to select the target soliton comb lines and nearby EO comb lines (highlighted in Fig.\,\ref{fig3}b) for optical-heterodyne detection. In the upper path, the lower-frequency soliton mode along with two adjacent EO comb modes yield a pair of beat notes at $\delta_{1}$\,$\approx$\,8.3\,GHz and $\delta_{2}$\,$\approx$\,7.7\,GHz respectively, with the constraint of $\delta_{1}+\delta_{2}=f_{\mathrm{rep,EO}}$. Likewise, another pair of beat notes are generated at $\delta_{3}$\,$\approx$\,6.1\,GHz and $\delta_{4}$\,$\approx$\,9.9\,GHz in the lower path, with $\delta_{3}+\delta_{4}=f_{\mathrm{rep,EO}}$. To identify the two comb modes by which a beat note is generated in each path, one can slightly tune the EO drive frequency ($f_{\mathrm{rep,EO}}$) while observing how the beat note shifts. For example, when $f_{\mathrm{rep,EO}}$ is increased, the beat note (at $\delta_{2}$ or $\delta_{3}$) formed by the soliton mode and the interior EO comb mode will shift to lower frequencies, as the interior EO comb mode comes closer to the soliton mode. Conversely, the other beat note (at $\delta_{1}$ or $\delta_{4}$) formed by the soliton mode and the exterior EO comb mode will shift to higher frequencies.

To access $f_{\mathrm{rep,Kerr}}$, we then mix two of the RF beat notes from different paths and record their frequency difference, $\delta=|\delta_{1}-\delta_{3}|$\,$\approx$\,2.2\,GHz. Fig.\,\ref{fig3}c shows a typical RF spectrum at the output of the RF mixer, whose input RF/LO frequency range is from 3.7 to 8.5\,GHz. Due to the limited conversion efficiency and finite isolation of the mixer, residual input RF beat notes can still been seen on the mixer output spectrum. They are shown here for illustration purpose, and can be further suppressed with RF filters. At last, we arrive at $f_{\mathrm{rep,Kerr}}=21 f_{\mathrm{rep,EO}}-\delta$, based on the measurement shown in Fig.\,\ref{fig3}. Note that the sign before $\delta$ should be reversed in the calculation if $\delta_{1}<\delta_{3}$.  

The electronic access of $f_{\mathrm{rep,Kerr}}$ allows us to investigate and quantify the tunability of $f_{\mathrm{rep,Kerr}}$ in LN microcombs. We vary the Kerr pump frequency and pump power separately while monitoring $\delta$ to infer the change in $f_{\mathrm{rep,Kerr}}$. Fig.\,\ref{fig4}a shows the measured $\delta$ beat note spectra as pump frequency is modulated. The $f_{\mathrm{rep,Kerr}}$-to-pump-frequency tuning rate is extracted to be 4.6\,MHz/GHz under an on-chip pump power of 50mW (Fig.\,\ref{fig4}b). Here, the pump laser frequency is piezo controlled by an arbitrary function generator and calibrated with a wavelength meter. On the other hand, one can also tune $f_{\mathrm{rep,Kerr}}$ by varying the pump power, e.g., via an external acousto-optic modulator (Fig.\,\ref{fig4}c). The $f_{\mathrm{rep,Kerr}}$-to-pump-power tuning rate is found to be -0.18\,MHz/mW around the 41.5-mW operating point (Fig.\,\ref{fig4}d).

We move on to characterize the frequency stability of the free-running $f_{\mathrm{rep,Kerr}}$. The drift of $f_{\mathrm{rep,Kerr}}$ is  reflected in $\delta$, since $f_{\mathrm{rep,EO}}$ is set by a reference RF synthesizer and has negligible frequency drift within the measurement time span (see Methods). Due to the limited amplitude of the $\delta$ beat note, we chose to use an electronic spectrum analyzer instead of a frequency counter \cite{Drake2019Terahertz,zhang2019terahertz,offchip_eo} to capture $\delta$ (see Methods). A time series measurement of $\delta$ is presented in Fig\,\ref{fig4}e, where the Kerr soliton comb is free-running. The corresponding Allan deviations (Fig\,\ref{fig4}f) reveal a fractional-frequency stability of 1.6\,$\times$\,$10^{-7}$ for 1-s measurements, which translates to a 53.4-kHz fluctuation in $f_{\mathrm{rep,Kerr}}$. The observed 1-s Allan deviation is 36 times higher than the previously reported value from a silica microtoroid of similar FSR \cite{zhang2019terahertz}. Several factors may contribute to the larger instability. First, LN is known to exhibit photorefactive effect which may introduce cavity resonance oscillations \cite{LNoscill}. Second, the Kerr soliton pump laser used here (TSL-710, Santec) has $>$\,10 times broader short-term linewidth than the laser used in \cite{zhang2019terahertz}. Furthermore, our fiber-to-chip edge coupling tends to be more sensitive to the environmental perturbations than the fiber-taper setup. The mitigation of these drifts will be subject of future investigations, for example, by minimizing the technical noise sources mentioned above and also by operating the Kerr soliton at a quiet point, as demonstrated in \cite{modecrossquiteRep,ultralowmicrowave}. 

For proof-of-concept stabilization of $f_{\mathrm{rep,Kerr}}$, we build a simple computer-aided feedback control loop (see Methods) to improve $f_{\mathrm{rep,Kerr}}$ frequency stability. We digitally sample $\delta$ and compare it to a frequency setpoint, and use the generated error signal to adjust the Kerr soliton pump laser frequency to compensate for the $f_{\mathrm{rep,Kerr}}$ drift. Due to the latency (60\,ms) of this digital approach, the effective feedback control bandwdith is limited to 17\,Hz. Fig.\,\ref{fig4}g shows the stabilized $f_{\mathrm{rep,Kerr}}$ with much reduced long-term drift compared with the free-running case (Fig.\,\ref{fig4}e). Allan deviations of the in-loop $f_{\mathrm{rep,Kerr}}$ also reveal better long-term stability, e.g., a two order of magnitude improvement is achieved for 100-s measurements (Fig.\,\ref{fig4}f). Additionally, the in-loop $f_{\mathrm{rep,Kerr}}$ frequency drift distribution is shown in Fig.\,\ref{fig4}h, exhibiting a standard deviation of 79\,kHz. For future improvement, we believe that a higher $\delta$ beat note SNR enabled by a stronger EO comb will permit the use of a faster frequency locker to improve both the long-term and short-term stability of $f_{\mathrm{rep,Kerr}}$ (see Methods).

\section{Discussion}
In summary, we demonstrate a monolithic, hybrid Kerr and EO combs platform based on a LN thin film which permits the electronic access and control of terahertz Kerr soliton repetition rates by leveraging the simultaneous Kerr and Pockels nonlinearities inherent to LN. The monolithic circuit integrates Kerr soliton and EO microcombs on a single chip, and provides an on-chip interface between fast Kerr solitons and readily detectable microwaves. We also point out opportunities for further improvement of this chip platform. For example, EO microring modulation efficiency can be enhanced by either accessing a larger LN EO tensor component $r_{33}$ which is 6 times higher than $r_{22}$ used here or incorporating a microwave resonator \cite{Zhang2019Broadband,resonantEOcomb}. The need for the auxiliary pump laser could also be relaxed by aligning a soliton mode to the target EO microring mode. The demonstrated monolithic scheme thus holds great promise for the future development of fully integrated optical frequency divider on the LN thin-film platform for optical clock application, as well as study of soliton dynamics in hybrid Kerr and EO systems. It should be also noted that the hybrid comb architecture demonstrated here can be extended to other materials platforms that possess simultaneous Kerr and Pockels nonlinearities, such as AlN \cite{AlNbeyondoctave}, AlGaAs \cite{submilliwatts}, and GaP \cite{Wilson2019Integrated}.

\vbox{}
\noindent \textbf{Methods}{\large\par}
\noindent \textbf{Device design.} We simulate microring dispersion with a finite-difference method solver (FIMMWAVE, Photon Design). To obtain suitable dispersion for Kerr soliton and EO microcomb generation, we adopt an air-clad partially etched waveguide structure, where the LN etching depth is 380\,nm leaving a 200-nm un-etched LN slab above $\mathrm{SiO_{2}}$. The Kerr soliton microring top-width and radius are designed to be 1.4\,$\mu$m, and 60\,$\mu$m respectively to yield a $\mathrm{D_{2}}/(2\pi)$\,$=$\,6\,MHz at 1550\,nm. The top width of the EO microring is set to be 1.95\,$\mu$m, giving rise to a low $D_{2}/(2\pi)$\,$=$\,2.5\,kHz. A 30\,$^{\circ}$ sidewall angle is included in the simulation to reflect the plasma etching-angle of the fabricated devices. The extracted microring dispersion profiles are presented in Supplementary Note 1. 

The Kerr soliton microring is constructed in an add-drop configuration. The drop-port waveguide directly guides solitons out of the chip, while the through-port waveguide couples to the following EO microring and delivers the hybrid combs for $f_{\mathrm{rep,Kerr}}$ detection. 

To favor Kerr soliton formation over SRS, the intracavity pump mode threshold energy for SRS ($\mathrm{\varepsilon_{th,R}}$) needs to be raised above the threshold for soliton generation ($\mathrm{\varepsilon_{th,S}}$) \cite{gong2020photonic,LNnearoctave}. In practice, we fix the drop-port coupling rate ($\mathrm{\kappa_{ex1,Kerr}}$) and adjust the through-port coupling rate ($\mathrm{\kappa_{ex2,Kerr}}$) to raise $\mathrm{\varepsilon_{th,R}}$ over $\mathrm{\varepsilon_{th,S}}$. The Kerr soliton microring $\mathrm{TE_{00}}$ modes exhibit a typical intrinsic decay rate of $\mathrm{\kappa_{in,Kerr}}/(2\pi)$\,$=$\,106\,MHz and $\mathrm{\kappa_{ex1,Kerr}}/(2\pi)$\,$=$\,13\,MHz (900-nm coupling gap) in the telecom band. For example, in order to generate a Kerr soliton with a 5.6-THz FWHM, $\mathrm{\kappa_{ex2,Kerr}/(2\pi)}$ should be made smaller than 72\,MHz. Here, we assume that $\mathrm{\kappa_{in,Kerr}}$, $\mathrm{\kappa_{ex1,Kerr}}$ and $\mathrm{\kappa_{ex2,Kerr}}$ are constant across the modes of interest, and the dominant Raman gain ($\mathrm{E(LO)_{8}}$) center \cite{gong2020photonic,LNnearoctave} overlaps a local soliton forming mode. In the experiment, we adopted a 850-nm through-port-Kerr-microring coupling gap, which yields a $\mathrm{\kappa_{ex2,Kerr}/(2\pi)}$\,$=$\,58\,MHz and leads to successful soliton generation (Fig.\,\ref{fig2}). SRS is also suppressed in the EO microring, since the intracavity pump mode energy is far below $\mathrm{\varepsilon_{th,R}}$. The EO microring exhibits a typical intrinsic decay rate of $\mathrm{\kappa_{in,EO}}/(2\pi)$\,$=$\,60\,MHz and a waveguide coupling rate of $\mathrm{\kappa_{ex,EO}/(2\pi)}$\,$=$\,20\,MHz (600-nm coupling gap). The measured $\mathrm{TE_{00}}$ resonance spectra of both microrings are shown in Supplementary Note 1. 

To allow for efficient interaction between the applied in-plane RF field and EO microring $\mathrm{TE_{00}}$ modes, we place 200-nm-thick GSG gold electrodes closely along the microring, keeping a 2\,$\mu$m gap between them (Fig.\,\ref{fig1}c). The $\mathrm{TE_{00}}$ resonance EO tuning efficiency (DC) is estimated to be 1.5\,pm/V based on COMSOL simulations and perturbation theory \cite{perturbTuning}. The simulation of EO microcomb generation and EO microring DC tuning characterization are detailed in Supplementary Notes 3 and 4 respectively.

\noindent \textbf{Nanofabrication.} The device fabrication started with a bare congruent LNOI wafer (NANOLN), which consists of a 600-nm-thick z-cut LN layer, a 2-$\mu$m-thick thermal $\mathrm{SiO_{2}}$ layer and a 500-$\mu$m-thick Si substrate. First, the LN layer was thinned down to the target 580-nm thickness by argon ($\mathrm{Ar^{+}}$) plasma etching \cite{LNnearoctave}. We then used electron beam lithography (EBL) to pattern the photonic circuits, and transferred the patterns into the LN layer with $\mathrm{Ar^{+}}$ plasma etching. The etching depth is about 380\,nm, leaving a 200-nm-thick LN slab unetched. Afterwards, we defined the electrode patterns via aligned EBL, deposited gold via thermal evaporation, and finished the fabrication with lift-off. 

\begin{figure}[h]
\centering
\includegraphics[width=0.9\linewidth]{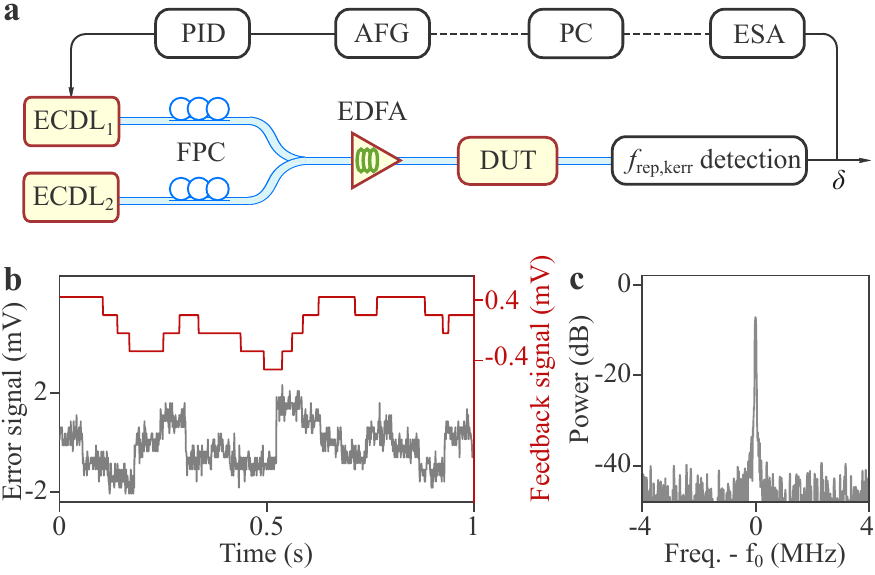}
\caption{\textbf{$f_{\mathrm{rep,Kerr}}$ feedback control}. \textbf{a} Schematic for stabilizing $f_{\mathrm{rep,Kerr}}$. ECDL, external cavity diode laser (TSL-710, Santec). $\mathrm{ECDL_{1(2)}}$ is used to generate the Kerr soliton (EO) microcomb; FPC, fiber polarization controller; EDFA, erbium-doped fiber amplifier; DUT, device under test (Fig.\,\ref{fig1}c); ESA, electronic spectrum analyzer; PC, personal computer; AFG, arbitrary function generator; PID, proportional-integral-derivative controller. The $f_{\mathrm{rep,Kerr}}$ detection is illustrated in Fig.\,\ref{fig3}a. $\delta$ is the target beat note. \textbf{b} A segment of the error signal (gray) output from the AFG, and the corresponding feedback signal (red) generated by the PID. \textbf{c} Zoom-in view of the $\delta$ beat note. $\mathrm{f_{0}}$\,$=$\,2.2146\,GHz.}
\label{fig_5}
\end{figure}

\noindent \textbf{Microwave drive and feedback control circuitry}. The 16-GHz microwave drive signal was generated by a commercial RF synthesizer (DS Instruments, SG24000H, phase noise -83\,dBc/Hz at 100\,Hz-10\,kHz offsets). The drive signal was amplified and passed through a 50-$\Omega$ terminated RF circulator before being delivered to the on-chip electrodes \cite{Zhang2019Broadband} via a GSG RF probe (ACP-40-A-GSG-200, FormFactor). 

The proof-of-concept slow-speed feedback loop for stabilizing $f_{\mathrm{rep,Kerr}}$ is shown in Fig.\,\ref{fig_5}a. We used an electronic spectrum analyzer (N9020a, Agilent) to capture $\delta$ with the built-in peak-search function. The recorded $\delta$ was then compared to a frequency setpoint on a computer. The frequency errors were converted to a voltage waveform by an arbitrary function generator (NI USB-6251). We fed the error signal to a proportional-integral-derivative controller (DigiLock 110, Toptica) to generate the feedback signal for adjusting the Kerr pump laser frequency via its internal piezoelectric transducer. 

Each feedback loop took approximately 60\,ms to complete, translating to an effective feedback bandwidth of 17\,Hz. While it took 9\,ms for the ESA to scan over a 8-MHz frequency span (1000 data points, 40-kHz RBW) and find $\delta$, most latency came from the computer executing the LabVIEW program (on Windows 10 operating system) and communication between the computer and instruments. Fig.\,\ref{fig_5}c shows one example of the generated error signal and the corresponding feedback signal. The error signal was proportional to the frequency error by a ratio of 0.1\,MHz/mV. The feedback signal drove the Kerr pump laser piezoelectric transducer, which has a frequency tunability of -3.3\,MHz/mV and a 1-kHz actuation bandwidth, to reduce $f_{\mathrm{rep,Kerr}}$ drifts. The on-chip Kerr pump power is fixed at 50\,mW during the process.

In the current system, the $\delta$ beat note exhibits an optimal SNR of 30\,dB at a 40-kHz RBW with the peak power reaching -17\,dBm (Fig.\,\ref{fig_5}b). The SNR is currently limited by the available EO comb sideband power (Fig.\,\ref{fig3}b). Since the EO comb is free-running during the measurement, cavity-pump detuning could drift and thus cause the EO comb line power to fluctuate. These factors combined prevent the fast locking of $f_{\mathrm{rep,Kerr}}$. We anticipate that the use of a microwave resonator \cite{Zhang2019Broadband,resonantEOcomb} and Pound-Drever-Hall stabilization technique \cite{thermal_nonlinear_dynamics} will improve the beat note SNR. A larger SNR, e.g., 53\,dB at a 40-kHz RBW, could allow using a frequency counter to directly read $\delta$, and a fast frequency locker (D2-135, Vescent Photonics, $>$500\,kHz locking bandwidth) to enhance both the short-term and long-term stability of $f_{\mathrm{rep,Kerr}}$.

\vbox{}
\noindent \textbf{Data availability} The data that support the findings of this study are available from the corresponding author upon reasonable request.

\noindent \textbf{Code availability} The codes used for simulations are available from the corresponding author upon reasonable request.

\def\bibsection{\section{\textbf{references}}}

\bibliographystyle{myaipnum4-1}

\vspace{1 mm}
\noindent \textbf{Acknowledgements.} This work was supported by DARPA under its ACES program. H.X.T acknowledges support from NSF Center for Quantum Networks (EEC-1941583). Funding for materials used for this study is partially provided by DOE (DE-SC0019406). The authors thank Yong Sun,  Sean Rinehart, Kelly Woods and Michael  Rooks for assistance in the device fabrication.

\vspace{1 mm}
\noindent \textbf{Author contributions.} H.X.T and Z.G. conceived the idea. Z.G. performed the device design, fabrication and measurement with the assistance from M.S., J.L, and J.S., Z.G. and M.S. performed the simulations. Z.G. and H.X.T. wrote the manuscript with the input from all other authors. H.X.T supervised the project.

\vspace{2 mm}
\noindent \textbf{Competing interests.} The authors declare no competing interests.

\vspace{2 mm}
\noindent \textbf{Additional information.} Supplementary information accompanies this manuscript.


%

\pagebreak
\widetext
\begin{center}
\textbf{\large Monolithic Kerr and electro-optic hybrid microcombs, Gong et al.}
\end{center}
\balancecolsandclearpage

\begin{figure}[h]
\centering
\includegraphics[width=\linewidth]{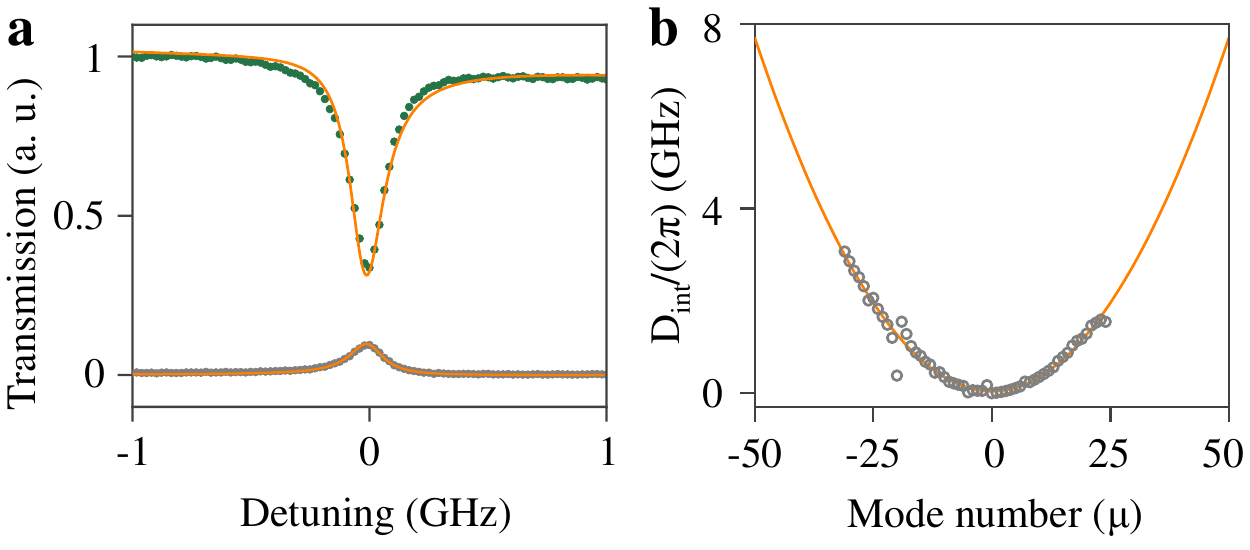}
\caption{Kerr soliton microring $\mathrm{TE_{00}}$ mode spectra and dispersion profile. \textbf{a} The pump mode spectra measured from the through-port (green dots) and drop-port (gray dots) waveguides, fitted with Fano functions (orange curve). The drop-port spectrum is normalized to the through-port transmission maximum. \textbf{b} Measured $\mathrm{TE_{00}}$ mode integrated dispersion (gray) around the pump mode at 1549.5\,nm, fitted with a quadratic function (orange).}
\label{figS1}
\end{figure}

Supplementary Figure \ref{figS1}\textbf{a} shows typical resonant spectra of the pump mode used for Kerr soliton generation, measured respectively from the through-port and drop-port waveguides. The dip in the through-port (peak in the drop-port) spectrum corresponds to a $\mathrm{TE_{00}}$ resonance at 1549.5\,nm. The mode intrinsic decay rate and external coupling rates can be extracted by fitting the spectra and found to be $\mathrm{\kappa_{in,Kerr}}/(2\pi)$\,$=$\,106\,MHz, $\mathrm{\kappa_{ex1,Kerr}}/(2\pi)$\,$=$\,13\,MHz (drop-port), and $\mathrm{\kappa_{ex2,Kerr}}/(2\pi)$\,$=$\,58\,MHz (through-port) respectively. The $\mathrm{TE_{00}}$ mode integrated dispersion is measured \cite{Dintmeas} and shown in Supplementary Figure \ref{figS1}\textbf{b}. Here, the dispersion measurement is limited by our laser scanning range (1480\,nm to 1640\,nm). A quadratic fitting reveals a dominant second order dispersion of $\mathrm{D_{2}}/(2\pi)$\,$=$\,6\,MHz \cite{Herr2013Temporal} over the telecomb band, agreeing well with the simulation presented in the main text.

A typical EO microring $\mathrm{TE_{00}}$ resonance spectrum at 1556\,nm is shown in Supplementary Figure \ref{figS2}\textbf{a}. The $\mathrm{TE_{00}}$ mode intrinsic decay rate and external coupling rate are extracted to be $\mathrm{\kappa_{in,EO}}/(2\pi)$\,$=$\,60\,MHz and $\mathrm{\kappa_{ex,EO}}/(2\pi)$\,$=$\,20\,MHz respectively. The measured integrated dispersion around the pump mode is presented in Supplementary Figure \ref{figS2}\textbf{b}. Strong mode-crossings are observed around $\mu$\,=\,-30, $\mu$\,=\,30 and $\mu$\,=\,90, causing local resonance frequency shifts which could distort EO comb spectrum. In particular, mode-crossings around $\mu$\,=\,30 and $\mu$\,=\,90 may be responsible for EO comb profile distortions at 193.4\,THz and 194.2\,THz respectively (see the second and fourth spectra in Fig. 2b of the main text). To mitigate mode-crossings, one could introduce a single-mode "filtering" section along the EO microring \cite{modefilter}.

\begin{figure}[h]
\centering
\includegraphics[width=\linewidth]{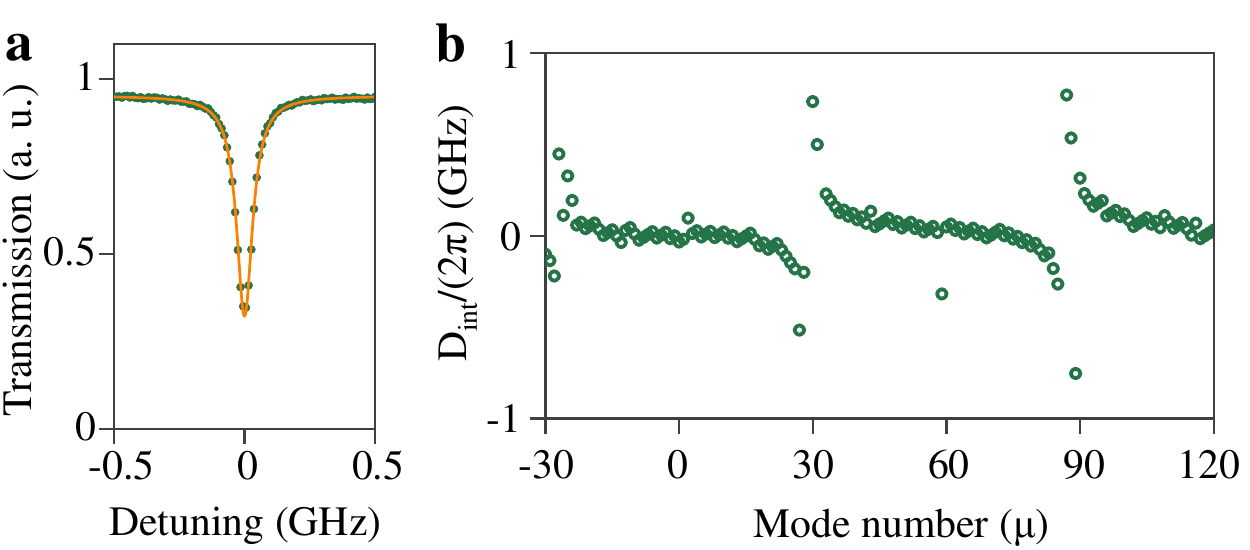}
\caption{EO microring $\mathrm{TE_{00}}$ mode spectrum and dispersion profile. \textbf{a} A $\mathrm{TE_{00}}$ resonance spectrum (green dots), fitted with a Fano function (orange curve). \textbf{b} Measured $\mathrm{TE_{00}}$ mode integrated dispersion around the pump mode at 1556\,nm.}
\label{figS2}
\end{figure}

The Mach-Zehnder interferometer (MZI) used for microring dispersion measurement \cite{Dintmeas} has a FSR of 50.2 MHz. Due to the relatively small EO microring dispersion ($D_{2}/2\pi$, estimated to be \,2.5\,kHz), we are unable to accurately extract the actual $D_{2}$ value within the vicinity of the pump mode before strong mode-crossings show up. We expect that reducing the MZI FSR, e.g., down to 200\,kHz, can help improve $D_{2}$ measurement accuracy.

\section*{Supplementary Note 2: \\Outgoing Kerr soliton microcomb from the drop-port waveguide}

\begin{figure}[h]
\centering
\includegraphics[width=\linewidth]{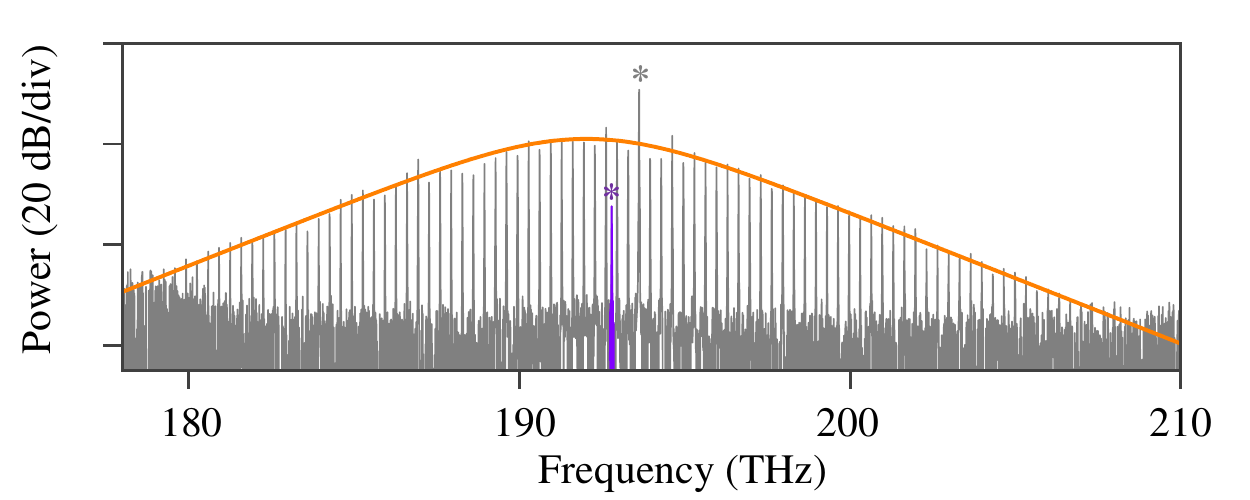}
\caption{A Kerr soliton microcomb spectrum measured from the drop-port waveguide (gray), fitted with a $sech^{2}$ function (orange). The Kerr soliton pump line is marked by the gray asterisk, and the EO pump through cross-talk is highlighted and marked by a purple asterisk.}
\label{figS3}
\end{figure}

The add-drop configuration of the Kerr soliton microring allows for direct access of the soliton microcomb from the drop-port waveguide, where the soliton is not filtered by the EO microring. Supplementary Figure \ref{figS3} shows one example of the outgoing Kerr soliton spectrum from the drop-port waveguide. The soliton spectrum exhibits a FWHM of 5.9\,THz under a 60-mW on-chip pump power. Here, a weak EO pump signal appears on the spectrum. This may be due to residual reflection in between facets and couplers.

\section*{Supplementary Note 3: \\Numerical simulations of electro-optic microcomb generation}

In the EO microring, the normalized electric field of a $\mathrm{TE_{00}}$ mode (by the sqaure root of its photon number) can be written as
\begin{eqnarray}
\vec{E}_{m}\left(\vec{r},t\right) & = & \vec{\phi}_{m}(\vec{r},\theta)e^{i\omega_{m}t}+\vec{\phi}_{m}^{*}(\vec{r},\theta)e^{-i\omega_{m}t},
\label{S01}
\end{eqnarray}
where $\phi_{j}(\vec{r},\theta)=\phi_{j}(\vec{r})e^{im\theta}$, $\omega_{m}$ and $m$ denote the spatial distribution, angular frequency and azimuthal number of the mode respectively. Optical modes are mutually orthogonal and their electric field distributions satisfy
\begin{align}
\begin{split}
\int\int drd\theta\varepsilon_{0}{\varepsilon_{r,m}}\left(\vec{r},\theta\right)\vec{\phi}_{m}(\vec{r},\theta)\cdot\vec{\phi}_{n}^{*}(\vec{r},\theta) \\ = \frac{1}{2}\hbar\omega_{m}\delta\left(m-n\right)\label{S02},
\end{split}
\end{align}
with $\varepsilon_{0(r)}$ referring to the vacuum (relative) permittivity. Now, consider that the EO microring is modulated by a strong in-plane RF electric field with an angular frequency of $\Omega=\omega_{m}-\omega_{n}$ and an amplitude distribution of $\vec{E}_{eo}(\vec{r})f(\theta)$ ($\vec{E}_{eo}$ is in units of V/m, representing the RF drive strength) respectively, the EO interaction Hamiltonian between the optical modes $m$ and $n$ can be given by 
\begin{align}
H_{nl} &= \frac{1}{2}\epsilon_{0}\int\int{\vec{E}}\cdot(\chi^{(2)}:\vec{E}\vec{E})drd\theta\nonumber \\
  &\approx  3\epsilon_{0}\int\int{r_{22}}E_{eo}\left(\vec{r}\right)f(\theta)\phi_{m}(\vec{r})\phi_{n}^{*}(\vec{r})e^{i(m-n)\theta}drd\theta a_{m}a_{n}^{\dagger}\nonumber \\
  & +h.c.
\label{S03}
\end{align}
where $\vec{E}$ is the total electric field of the two optical modes and the RF drive. $a_{m}$ and $a_{n}$ are the Bosonic operators of the two optical modes. Here, we neglect high-frequency rotating terms
$a_{m}a_{n}+h.c.$ and only consider the dominant electric field components of the $\mathrm{TE_{00}}$ modes and the RF drive (along LN Y-axis). Based on Eq.\,\ref{S02} and \ref{S03}, we introduce the EO coupling strength ($g_{\mathrm{EO}}$) as
\begin{widetext}
\begin{align}
\hbar g_{\mathrm{EO}} = & 3\epsilon_{0}\int\int{r_{22}}E_{eo}\left(\vec{r}\right)f(\theta)\phi_{m}(\vec{r})\phi_{n}^{*}(\vec{r})e^{i(m-n)\theta}drd\theta\nonumber \\
= & 3\epsilon_{0}{r_{22}}\sqrt{\frac{\hbar\omega_{m}\hbar\omega_{n}}{4\epsilon_{0}\epsilon_{r,m}\epsilon_{0}\epsilon_{r,n}}}\times\frac{\int\int{E_{eo}\left(\vec{r}\right)} f(\theta)\phi_{m}(\vec{r})\phi_{n}^{*}(\vec{r})e^{i(m-n)\theta}drd\theta}{\sqrt{\int\int|\phi_{m}(\vec{r})|^{2}d\vec{r}d{\theta}}\sqrt{\int\int|\phi_{n}(\vec{r})|^{2}d\vec{r}d{\theta}}}.
\label{S04}
\end{align}
\end{widetext}
Note that in order to achieve effective coupling between the two optical modes, the RF field azimuthal distribution $f(\theta)$ needs to compensate the phase mismatch between optical modes, i.e., $e^{i(m-n)\theta}$.

In the experiment, the RF drive is strong and its frequency is set close to the EO microring FSR, leading to an effective linear coupling between adjacent $\mathrm{TE_{00}}$ modes under non-depletion approximation \cite{resonantEOcomb}.
The linearised interaction Hamiltonian among $\mathrm{TE_{00}}$ modes can be written as
\begin{equation}
H_{\mathrm{EO}}=\frac{1}{2}g_{\mathrm{EO}}\sum_{j}(a_{j}a_{j+1}^{\dagger}+a_{j}a_{j-1}^{\dagger})+h.c..\label{S05}
\end{equation}
Assuming that $\int|\phi_{j}(\vec{r})|^{2}d\vec{r}$ is constant across the $\mathrm{TE_{00}}$ modes of interest and the RF drive has a 14-V peak amplitude, $g_{\mathrm{EO}}/(2\pi)$ can be estimated as 280\,MHz via COMSOL simulations.

Based on the system Hamiltonian $H$\,$=$\,$H_{0}$\,$+$\,$H_{EO}$ ($H_{0}$\,$=$\,$\sum_{j}\hbar\omega_{j}a^{\dagger}_{j}a_{j}$, ignoring Kerr and Raman effects), the mean field dynamics of each $\mathrm{TE_{00}}$ mode can be derived as follows
\begin{align}
\frac{d}{dt}a_{\mu} = & -(\kappa_{\mu}/2+i\Delta_{\mu})a_{\mu}-\frac{i}{2}g_{\mathrm{EO}}\sum_{k}[a_{k}\delta\left(k-\mu-1\right)\nonumber\\  & +a_{k}\delta\left(k-\mu+1\right)].\label{S06}
\end{align}
where $\Delta_{\mu}=\mathrm{D}_{\mathrm{int},\mu}+\omega_{0}-\omega_{\mathrm{P}}+\mu(D_1-\Omega)$, with $\omega_{0(\mathrm{P})}$ denoting the pump resonance (pump field) angular frequency, and $D_{\mathrm{1}}/(2\pi)$ representing the EO microring FSR. $\kappa_{\mu}$ is the total decay rate for the $\mu^{\mathrm{th}}$ $\mathrm{TE_{00}}$ mode, consisting of intrinsic cavity decay rate ($\kappa_{\mathrm{i},\mu}$) and coupling rate to the external bus waveguide ($\kappa_{\mathrm{e},\mu}$). The $\delta$ functions in Eq.\,\ref{S06} accounts for the phase-matching condition.

We move on to simulate EO microcomb formation by numerically solving the coupled mode equations (Eq.\,\ref{S06}) using experimental parameters. The simulated spectrum is shown in Supplementary Figure\,\ref{figS4}, which agrees well with measured spectrum. 

\begin{figure}[h]
\centering
\includegraphics[width=\linewidth]{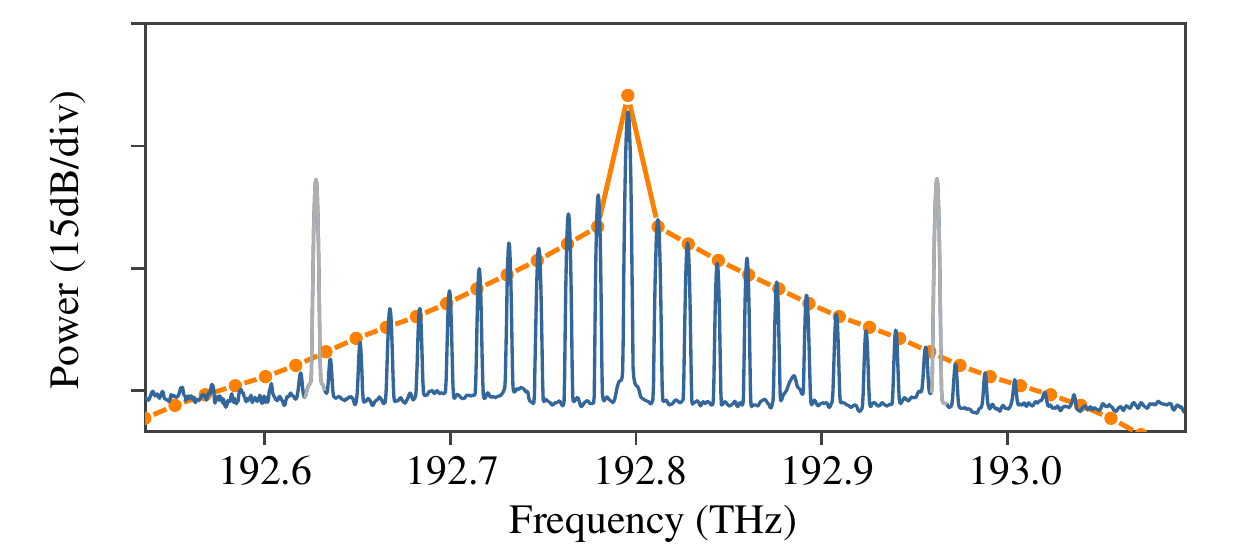}
\caption{Measured (blue) and simulated (orange dot-line) EO microcomb spectra. The measured spectrum is the same as in Fig.\,3(b) of the main text. Note that the two neighboring Kerr soliton comb lines (gray) are also present on the spectrum. The orange dots represent the simulated EO comb line power.}
\label{figS4}
\end{figure}

\section*{Supplementary Note 4: \\Characterization of EO microring DC tuning}

Finally, we experimentally and numerically characterized the modulator DC tuning efficiency. To avoid cancellation of EO modulation (a GSG probe was used here), we applied DC voltages across one racetrack arm and measured $\mathrm{TE_{00}}$ resonance spectra. The shifted $\mathrm{TE_{00}}$ resonance spectra under different DC offsets are plotted in Supplementary Figure \ref{figS5}\textbf{a}. The resonance tuning efficiency is extracted to be 1\,pm/V for modulating one racetrack arm, which infers a $\sim$\,2-pm/V tuning efficiency for modulating the entire microring in a push-pull manner \cite{LNmodulators}. The inferred tuning efficiency is a bit higher than the simulated one (1.5\,pm/V). The difference may be due to the mode crossing \cite{modecrossingTuning} and uncertainties in material parameters. In simulations, LN refractive indices along Y(X)-axis and Z-axis are set as 2.21 and 2.14 respectively; LN DC dielectric constants along Y(X)-axis and Z-axis are set as 80 and 30 respectively.  

\begin{figure}[h]
\centering
\includegraphics[width=\linewidth]{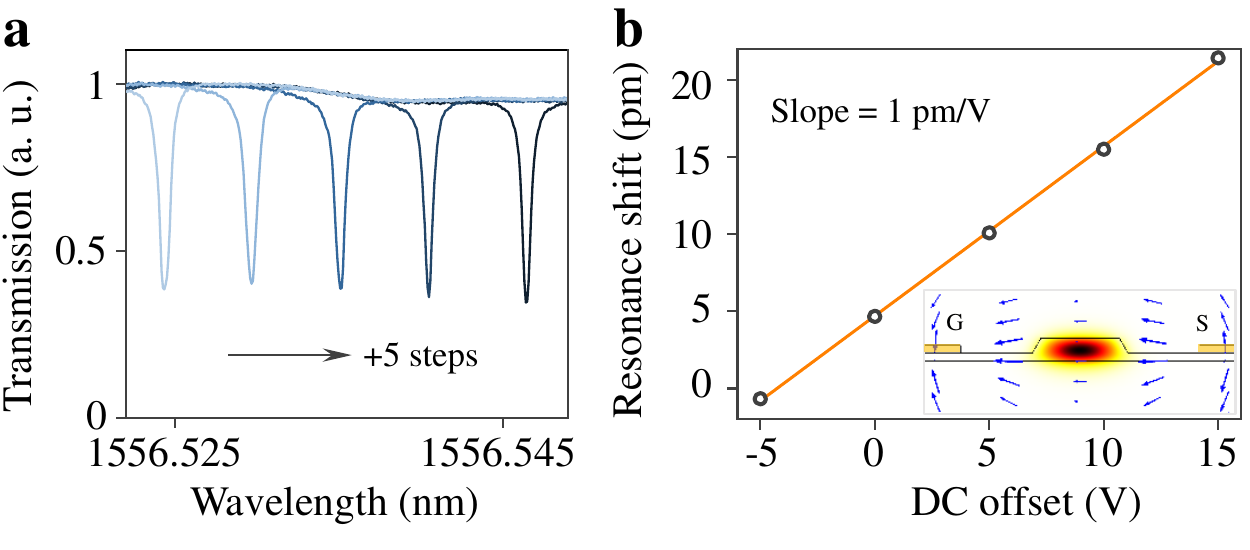}
\caption{DC electro-optic tuning of the EO microring. \textbf{a} Shifted $\mathrm{TE_{00}}$ resonance spectra under different DC offsets. The voltages were applied with a 5-V incremental step. \textbf{b} Measured $\mathrm{TE_{00}}$ resonance shifts versus applied DC offsets (circle) along with a linear fitting curve (yellow). Inset: simulated $\mathrm{TE_{00}}$ mode profile (heat map) and DC electric field (blue arrows) on top of the device cross-section. The gold electrodes are colored yellow; black lines delineate LN waveguide profile. Voltages are applied between the signal (S) and ground (G) electrodes.}
\label{figS5}
\end{figure}

\end{document}